# Physics and measurements of magnetic materials

*S. Sgobba*
CERN, Geneva, Switzerland

**Abstract**
Magnetic materials, both hard and soft, are used extensively in several components of particle accelerators. Magnetically soft iron–nickel alloys are used as shields for the vacuum chambers of accelerator injection and extraction septa; Fe-based material is widely employed for cores of accelerator and experiment magnets; soft spinel ferrites are used in collimators to damp trapped modes; innovative materials such as amorphous or nanocrystalline core materials are envisaged in transformers for high-frequency polyphase resonant convertors for application to the International Linear Collider (ILC). In the field of fusion, for induction cores of the linac of heavy-ion inertial fusion energy accelerators, based on induction accelerators requiring some $10^7$ kg of magnetic materials, nanocrystalline materials would show the best performance in terms of core losses for magnetization rates as high as $10^5$ T/s to $10^7$ T/s. After a review of the magnetic properties of materials and the different types of magnetic behaviour, this paper deals with metallurgical aspects of magnetism. The influence of the metallurgy and metalworking processes of materials on their microstructure and magnetic properties is studied for different categories of soft magnetic materials relevant for accelerator technology. Their metallurgy is extensively treated. Innovative materials such as iron powder core materials, amorphous and nanocrystalline materials are also studied. A section considers the measurement, both destructive and non-destructive, of magnetic properties. Finally, a section discusses magnetic lag effects.

## 1   Magnetic properties of materials: types of magnetic behaviour

The sense of the word 'lodestone' (waystone) as magnetic oxide of iron (magnetite, $Fe_3O_4$) is from 1515, while the old name 'lodestar' for the pole star, as the star leading the way in navigation, is from 1374. Both words are based on the original 'lode' spelling of 'load', issued from the old (1225) English 'lad', guide, way, course [1]. According to tradition, the mariner Flavio Gioia of Amalfi, born 1302, first discovered the 'power of the lodestone' enabling the manufacture of the first compass and replacing the lodestar in navigation. Nevertheless magnetite, known according to tradition to the Chinese since 2600 B.C., is cited first in Europe by Homer, relating that lodestone was already used by the Greeks to direct navigation at the time of the siege of Troy [2].

Magnetic properties of several materials are discussed in the text by Bozorth [3]. Conventional soft and hard magnetic materials are treated in Ref. [4]. The volume of O'Handley [5] covers a number of advanced materials, including amorphous and nanocrystalline materials. A general introduction to magnetic properties of materials can be found in the recent textbook by Cullity and Graham [6]. The comprehensive *Handbook of Magnetism and Advanced Magnetic Materials* [7] systematically covers very novel materials of technological and scientific interest in volume 4, including advanced soft magnetic materials for power applications.

Diamagnetism is due to induced currents opposing an applied field resulting in a small negative magnetic susceptibility $\kappa$. Diamagnetic contributions are present in all atoms, but are generally negligible in technical materials, except superconducting materials under some conditions. Monoatomic rare gases such He are diamagnetic, as well as most polyatomic gases such as $N_2$ (that might show nevertheless a net paramagnetic behaviour because of $O_2$ contamination). Since He is repelled by magnetic fields, operation of superconducting magnets in a weightless environment during orbital flights imposes a significant difficulty not present in laboratory experiments, already discussed and quantified in 1977 [8]. This concern is still present today: the effect of a magnetic field on diamagnetic liquid helium will be studied in the very near future in the cryogenic system of the cryomagnet of the Alpha Magnetic Spectrometer (AMS) experiment, foreseen on the International Space Station (ISS) [9].

Paramagnetism, corresponding to a positive susceptibility, is observed in many metals and substances including ferromagnetic and antiferromagnetic materials above their Curie ($T_c$) and Néel ($T_N$) temperature, respectively [10]. Particular care should be taken for some Ni-basis superalloys for non-magnetic application at very low $T$. Incoloy 800 (32.5Ni-21Cr-46Fe) features a magnetic permeability as low as 1.0092 at room temperature (annealed state, under a field of 15.9 kA/m). Nevertheless, due to a $T_c$ = -115 °C, the alloy is ferromagnetic at cryogenic temperatures.

Ferromagnetism is due to the ordered array of magnetic moments, caused by the interaction of atomic spin moments occurring in certain conditions. Field-dependent permeability and persistent magnetization after the removal of magnetic field are observed for hysteretic ferromagnetic materials. Ordered ferromagnetic phase occurs for ferromagnets at $T < T_c$. Here $T_c$ is the temperature above which spontaneous magnetization 'vanishes' [6]. The $T_c$ of Fe, Ni and Co are 1043 K, 631 K and 1394 K, respectively. In general, ferrous alloys with body centred cubic (bcc) crystalline structure are ferromagnetic, while face centred cubic (fcc) are not. Nevertheless, rapidly solidified metastable alloys such as Fe-Cu alloys can show ferromagnetism in a wider composition range than expected, even in the fcc phase formed below 70% Fe content [11].

Antiferromagnetism corresponds to an antiparallel arrangement with zero net magnetic moment at $T < T_N$. 'Non-magnetic' austenitic stainless steels such as AISI 304L, 316L, 316LN, high Mn – high N stainless steels are antiferromagnetic under $T_N$ and paramagnetic above $T_N$, where they obey a Curie–Weiss law ($\kappa = C/(T-\Theta)$), where $\Theta$ is a negative critical temperature and $C$ is a constant (Fig. 1a).

Magnetic susceptibility of high Mn – high N grades such as P506 (approx. 0.012%C, 19%Cr, 11%Ni, 12%Mn, 0.9%Mo, 0.33%N) and UNS 21904 (approx. 0.028%C, 20%Cr, 7%Ni, 9%Mn, 0.35%N), particularly at 4.2 K, is lower than any traditional steel of the 300-series. As known, this is essentially due to the higher Mn content of the alloys (P506, Mn = 12%; UNS21904, Mn = 9%), stabilizing austenite (fcc 'non-magnetic' phase), and increasing $T_N$. Higher $T_N$ allows for lower values of $\kappa$ ($<3\cdot10^{-3}$) at 4.2 K. Measured values of $T_N$ are in agreement with the Warnes [12] law:

$$T_N/K = 90 - 1.25\text{Cr} - 2.75\text{Ni} - 5.5\text{Mo} - 14\text{Si} + 7.75\text{Mn} \quad . \tag{1}$$

As an example, for steel P506, predicted $T_N$ = 121.5 K, measured $T_N$ is 125.7 K. Owing to the absence of precipitated δ-ferrite (bcc magnetic phase) in the weld, the presence of a laser weld has no measurable influence on the magnetic susceptibility of P506 (Fig. 1b). On the other hand, in welds of UNS21904, δ-ferrite contributes a significant increase of susceptibility in the whole $T$ range [13].

Diamagnetism and paramagnetism can be considered as mainly due to the magnetic contribution of isolated atoms or molecules (in reality the existence of a Curie temperature $T_c$ is explained by interaction of elementary moments in the paramagnetic range). Ferromagnetism and antiferromagnetism are due to a larger order arrangement of electron spins and/or magnetic moments.

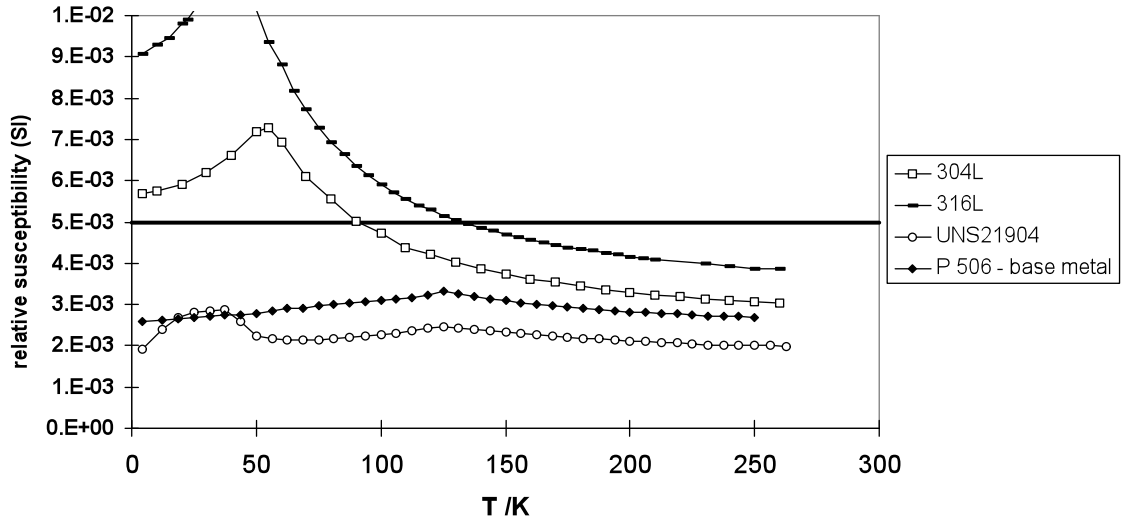

**Fig. 1a:** Magnetic susceptibility of different steels of the AISI 300 series, compared to high Mn – high N steels P506 and UNS 21904. Maximum allowed limit at CERN for non-magnetic applications is $5 \cdot 10^{-3}$. Peaks are at the respective Néel temperatures $T_N$. Above $T_N$, susceptibility obeys a Curie–Weiss law $\kappa = C/(T-\Theta)$.

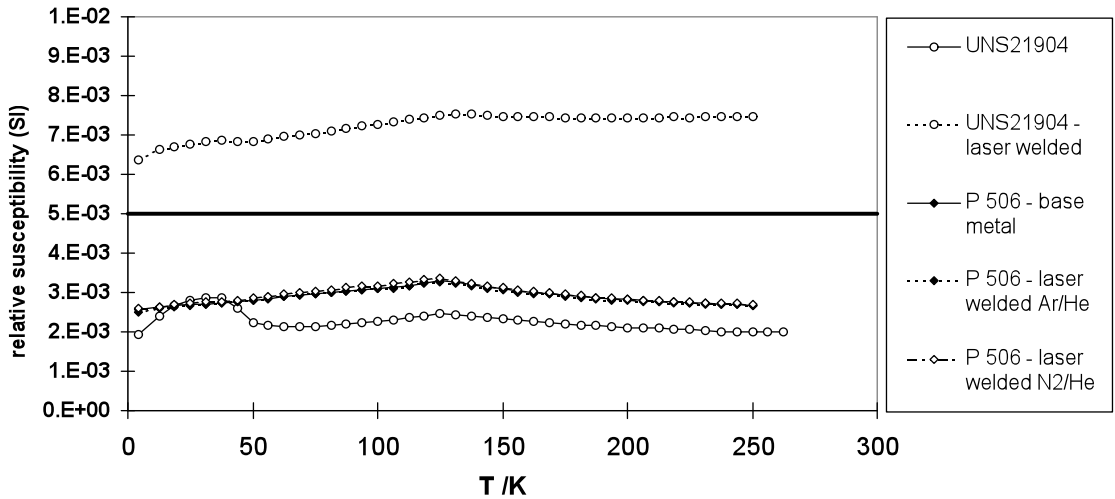

**Fig. 1b:** Compared magnetic susceptibility of steels P506 (base metal and weld) and UNS 21904 (idem). Maximum allowed limit at CERN in the welds for non-magnetic applications is $5 \cdot 10^{-3}$.

## 2 Soft ferromagnetic materials of interest for accelerator technology

### 2.1 Some definitions and units

Working in SI, we define the flux density or magnetic induction $B$ (measured in T) and the magnetic field strength $H$ (A m$^{-1}$). The permeability $\mu$ (H m$^{-1}$) is defined by

$$B = \mu \cdot H \tag{2}$$

The magnetization $M$ (A m$^{-1}$) is defined as

$$B = \mu_0 \cdot (H + M) \tag{3}$$

where $\mu_0$ (H m$^{-1}$) is the permeability of free space. The susceptibility $\kappa$ (dimensionless) is the ratio $M/H$. From the above

$$\mu = \mu_0 \cdot (1 + \kappa) \tag{4}$$

The relative permeability is defined as $\mu_r = \mu / \mu_0$. From the above relationships, $\mu_r = 1 + \kappa$. A relative permeability of 1.005 corresponds to a susceptibility of $5 \cdot 10^{-3}$. The relative permeability $\mu_r$ and susceptibility $\kappa$ are material properties, frequently reported for both magnetic and 'non-magnetic' materials.

### 2.2 Magnetization curves of soft ferromagnetic materials

A magnetization curve is the plot of the intensity of magnetization $M$ or the magnetic induction $B$ against the field strength $H$.

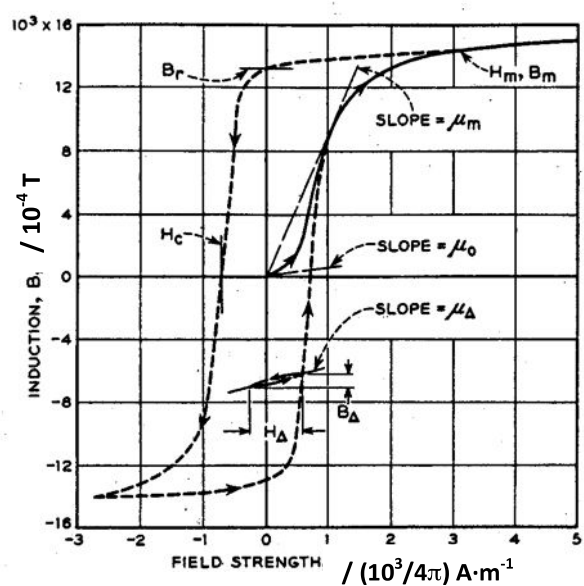

**Fig. 2:** Magnetization curve and hysteresis loop of iron (from Bozorth [3])

In the example of Fig. 2, the values of the field strength $H_m$ and the magnetic induction $B_m$ at the tip of the loop are defined. In the hysteresis loop, are also defined the residual induction $B_r$ for which $H = 0$ (called retentivity if the tip corresponds to saturation) and the coercive force $H_c$ for which $B = 0$ (called coercivity if the tip corresponds to saturation). For ferromagnetic materials, permeability is strongly dependent on field and tends to 1 for saturation. The initial and maximum permeability are easily identified in the curve. The magnetic properties of ferromagnetic materials are significantly affected by their purity, the metalworking processes applied to the material (hot and cold working, subsequent annealing), and the resulting microstructure. Anisotropy effects due to texture can occur (effect of rolling, extrusion). By definition, soft ferromagnetic materials are easily magnetized and demagnetized (materials for transformer cores, for shielding of magnetic fields, magnetically soft ferrites for ac shielding applications, etc.). They have narrow hysteresis loops (low values of $H_c$), high permeability, low eddy-current losses, high magnetic saturation inductions.

## 2.3  High-purity iron

Iron is referred to as 'high purity' when the total concentration of impurities (mainly C, N, O, P, S, Si and Al) does not exceed a few hundred ppm. Otherwise it is rather referred to as low carbon steel or non-alloyed steel [14]. Very pure Fe features a high electrical conductivity and is unsuitable for ac applications. Typical impurity contents of different grades of iron are shown in Table 1.

**Table 1:** Impurity content of different iron grades. So-called 'Armco irons' can correspond to very different purity grades (from Ref. [15]).

| Element | Concentration (at.%)×10⁶ High-purity iron | ORNL Armco iron | BMI Armco iron |
|---|---|---|---|
| Al | 0.21–2.1 | <105 | <0.4 |
| Ca | 0.14–1.4 | | |
| Cr | | <54 | <0.9 |
| Cu | 0.09–0.9 | 90 | 0.09–0.9 |
| Mn | | 51 | 1–10 |
| Mo | | <29 | <6 |
| Ni | 0.95–9.5 | 95 | 1–10 |
| Si | 2.0–20 | <40 | 0.2–2 |
| Ti | | <12 | <1.1 |
| V | | <22 | <1.1 |
| C | 14 | 61 | 75 |
| P | 2.0 | 11 | 16 |
| S | 5.2 | 40 | 51 |
| H₂ | <5.6 | <5.6 | 95 |
| O₂ | 8.8 | 304 | 210 |
| N₂ | 2.0 | 20 | 328 |
| **Totals:** | | | |
| Minimum including oxygen | 40.99 | 672 | 777 |
| Minimum without oxygen | 32.19 | 368 | 567 |
| Maximum including oxygen | 71.50 | 940 | 807 |
| Maximum without oxygen | 62.7 | 636 | 597 |

ᵃ Al, Ca, Cr, Cu, Mn, Mo, Ni, Si, Ti, V analyzed by emission spectroscopy (semiquantitative); C, P, S, H₂, O₂, N₂ analyzed by quantitative analysis.

Table 2 summarizes magnetic properties of various grades of iron. Saturation magnetization (≈ 2.15 T) is not strongly influenced by purity, while coercivity $H_c$ and achievable magnetic permeability do strongly depend on purity and crystallographic features. Values of initial and maximum permeability drop for cold worked material. In order to restore magnetic properties, annealing cycles are required, allowing internal strains to be reduced, grain size to be increased, as well as the annealing of dislocations. Iron has various phases with different stability domains: α- and δ-iron, corresponding to the ferromagnetic ferritic phase of bcc structure, which are present up to

912°C ($T_{\alpha,\gamma}$) and in the ranges between 1394°C and 1538°C, respectively, and γ-iron, corresponding to the austenitic phase of fcc structure, in the range between 912°C and 1394°C. This phase is non-ferromagnetic. For this reason, there are two classes of anneals used commercially [16]:

1) Anneals below 900°C

2) Anneals at or about 925°C or higher to promote grain growth and to further improve magnetic properties.

These anneals, particularly the high $T$ ones, should be followed by slow cool. Higher maximum permeability is obtained by exceeding $T_{\alpha,\gamma}$, allowing the material to enter the γ stability domain and subsequently revert γ by slow cooling. For high maximum permeability, annealing should be performed at between 925°C and 1000°C (above $T_{\alpha,\gamma}$) followed by cooling at a rate < 5°C/min. For high permeability at $B \geq 1.2$ T, it is advisable to anneal at a maximum of 800°C and to cool slowly [17].

**Table 2:** Magnetic properties of various grades of iron (from Ref. [14]).

| Material | $H_c$ (A m$^{-1}$) | $\mu_i$ ($\mu_0$) | $\mu_{max}$ ($\mu_0$) |
|---|---|---|---|
| Ingot (99.8% Fe) | 112 | 10 | 1 000 |
| Armco | 80 | 200 | 7 000 |
| Commercially pure | 20–100 | 200–500 | 3 500–20 000 |
| Carbonyl iron powder | 6 | 3 000 | 20 000 |
| Vacuum-melted | 25 | — | 21 000 |
| Electrolytic | 7 | 1 000 | 26 000 |
| Electrolytic annealed | 18 | — | 41 500 |
| Vacuum-smelted and hydrogen-annealed | 3 | — | 88 400 |
| Purified Armco (99.95% Fe) | 4 | 10 000 | 227 000 |
| Vacuum-annealed | — | 14 000 | 280 000 |
| Single-crystal | — | — | 680 000 |
| Single-crystal, magnetically annealed | 12 | — | 1 430 000 |

Saturation magnetic polarization $J_s = 2.15$ T at 20°C, except for ingot with $J_s = 2.05$ T.

## 2.4 Low-carbon steels

For applications that require 'less than superior magnetic properties' [4], low-C steels are frequently used, including in magnet construction where in several cases they are purchased to magnetic specification. One of the most common grades is the structural–constructional steel 1010[1] used for the VINCY cyclotron, in the OPERA and ATLAS experiments, and recently proposed for the CLIC main beam quadrupole prototypes. Owing to large ranges and allowed impurity contents, the composition of a specific grade of low-C steel is not sufficiently reproducible between different producers and heats to closely guarantee magnetic properties. For the 1010 steel, Si varies in different possible content ranges depending even on the form of the product (bars, rods, etc.). This explains the large spread in magnetization curves for different heats of the same grade of steel (Fig. 3). The recommended magnetic annealing cycle for this steel is at 815°C ≤ $T$ ≤ 980°C, for a duration of between 1 h and 6 h followed by furnace cooling [4]. As for pure Fe, lower values of $T$ are intended for stress relief, and higher range for full annealing. During cooling, particularly in the critical temperature range (between 849°C and 682°C), slower cooling rates should be applied (for 1010 steel at a rate of 28°C/h).

Contrary to high-purity irons (for C concentrations less than 20 ppm), low-C steels are subject to magnetic ageing. An increase of coercivity occurs with time, due to formation of cementite precipitates giving rise to domain wall pinning. For magnetic cores that may operate at between 50°C and 100°C, ageing can be an issue. Ageing is mainly due to C, but also to N (due to formation of AlN precipitates), and S. In order to avoid ageing, C, N, S should be reduced below the range 20 ppm to 30 ppm. This reduction is possible by degassing of the melt followed by a final purification of the steel under pure hydrogen, at a $T$ as high as 1475°C. Indeed, at high $T$, $H_2$ reacts with the C present in the steel through the reaction $Fe + C + 2H_2 = Fe + CH_4$ and with other impurities such as oxygen, sulfur, nitrogen by forming $H_2O$, $H_2S$, $NH_3$, respectively, thus reducing their content under critical concentrations.

---
[1] Roughly equivalent to European grades 1.1121, 1.0301, 1.0308, 1.0032.

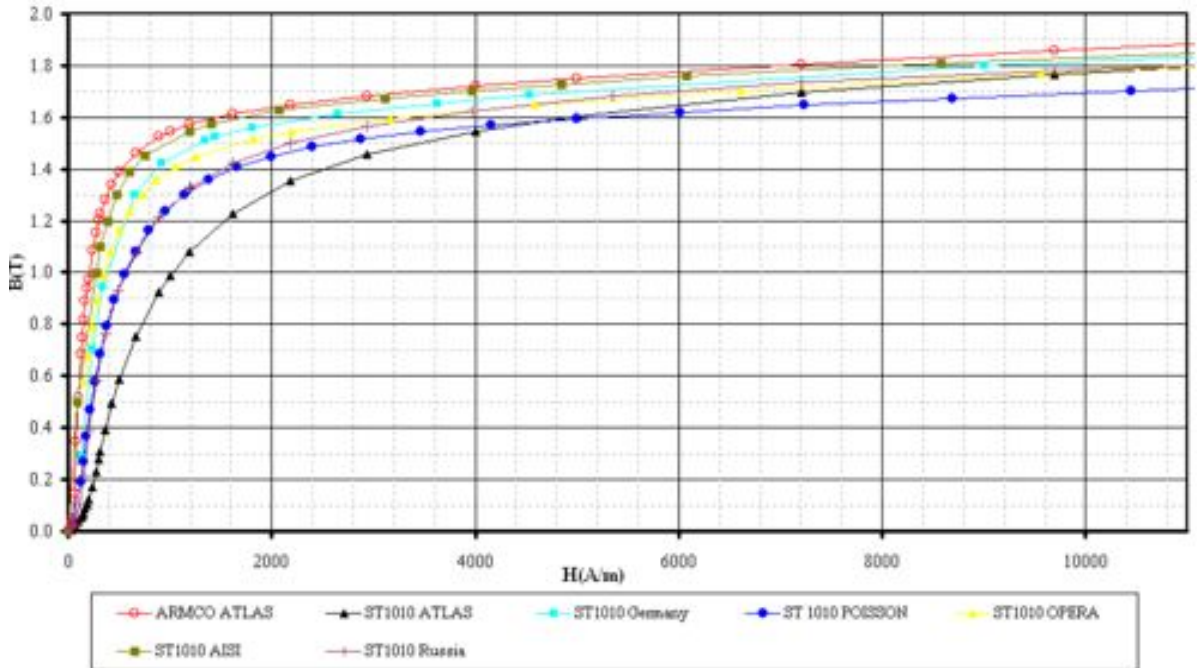

**Fig. 3:** Magnetization curves of different heats of 1010 constructional steels used as magnetic steel in different experiments (courtesy A. Vorozhtsov)

## 2.5  Non-oriented silicon steels

The Fe-Si alloys (phase diagram in Fig. 4) were accidentally discovered by Hadfield in 1882 (1.5% Si content). Magnetic properties were reported in 1900. Properties of earlier alloys (Hopkinson, 1885) were hindered by an excessive C content. Industrial production started in Germany in 1903 (2–2.5% Si). Their commercial use started in the US in 1905 and in England in 1906 [3].

Alloying with Si allows for an increase in permeability and decrease in hysteresis loss. Also thanks to additions of Al and Mn, eddy current losses decrease due to higher resistivity. With Al addition that reacts with N to form AlN, no ageing is experienced by silicon steels. On the other hand, with respect to pure Fe, saturation magnetization decreases with increasing Si content (2 T for 3.5% Si).

Silicon steels, also called electrical steels, are industrially produced in casting-hot rolling lines. Hot-rolled strips are subsequently pickled, cold rolled, continuous annealed (annealed products are called 'fully processed'), coated and slitted on line to the required width [19]. Coatings play an important role for the adhesive bonding of magnet laminations. They provide electrical insulation in addition to mechanical bonding. Coatings are designated according to the IEC 60404-1-1 standard [20], can be organic, inorganic with organic components, can be applied on one or both sides. Advanced coatings based on active organic bonder lacquers such as STABOLIT 70® require a delicate curing operation at a later stage, after stacking of the laminations. This coating can reach a mean shear strength above 22 MPa when cured at 190°C for 15 min [21], which stays above 20 MPa for doses up to $10^8$ rad. Fully processed non-oriented silicon steels can be easily specified according to EN 10106 [22].

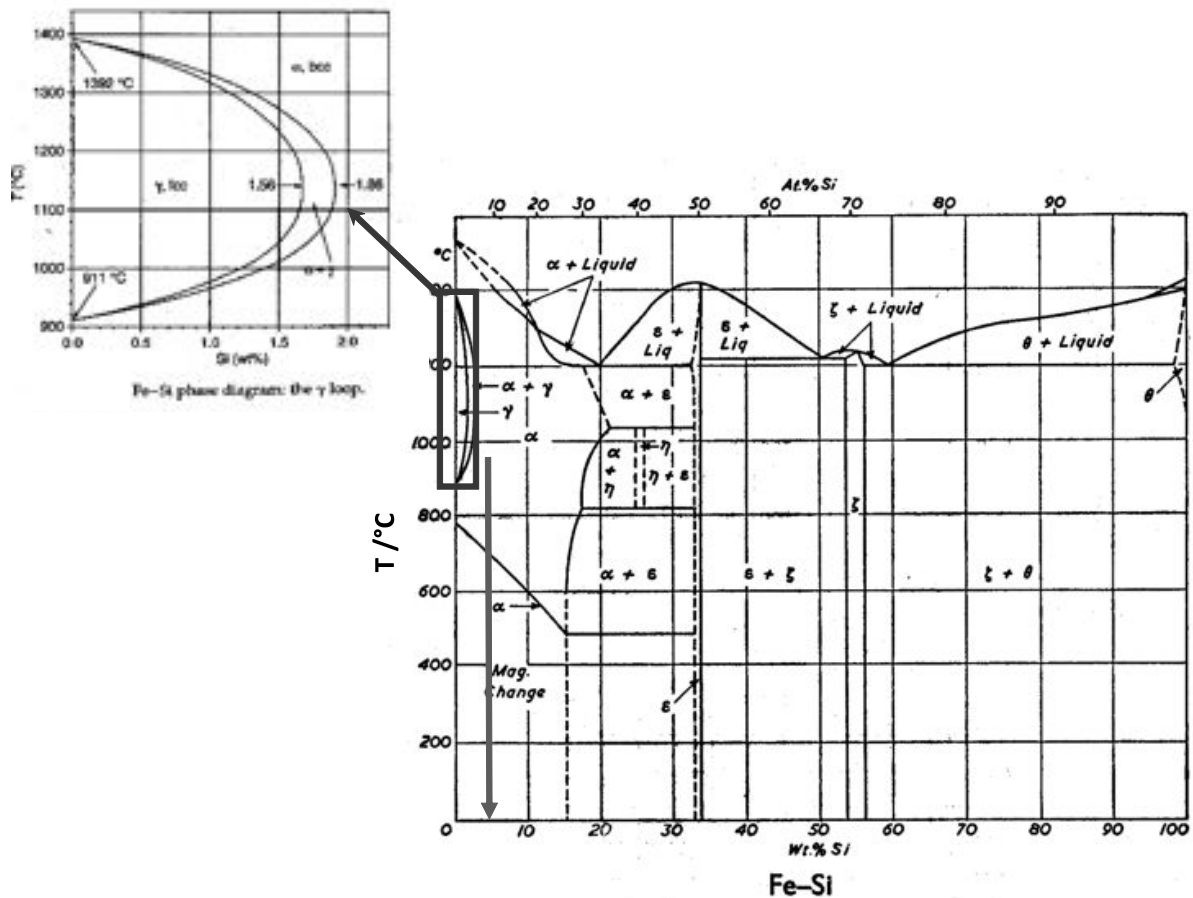

**Fig. 4:** Phase diagram of the Fe-Si system (from Ref. [18]) and detail of the austenitic 'γ loop' (from Ref. [14]). Si content of 3.5% (vertical arrow) represents an upper industrial limit (limited ductility for higher contents).

## 2.6 Oriented silicon steels

Iron single crystals exhibit minimum coercivity and maximum permeability when magnetized along one of the <001> axes. Fe-Si is also most easily magnetized in this direction. The so-called 'Goss' texture (110)[001] (Fig. 5) can be developed in silicon steels by a controlled sequence of cold rolling and annealing steps.

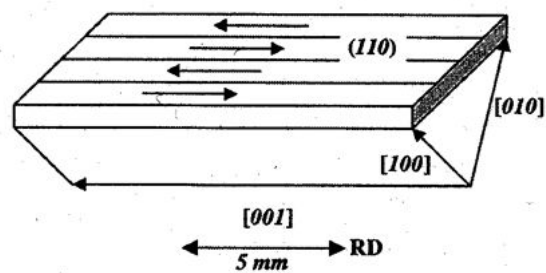

**Fig. 5:** Prevalent orientation of grain crystal axes with respect to rolling direction (RD) in grain oriented sheets of silicon steels (from Ref. [14])

Control of the texture, achievement of a large grain size and low impurity content will allow coercive fields as low as 4 A/m to 10 A/m and maximum permeability around $5 \times 10^4$ to be achieved in grain-oriented (GO) alloys. These figures are approximately 10 times higher than in non-grain-oriented (NGO) silicon steels. Conventional GO (7° dispersion of the [001] axes around RD) sheets represent 80% of the market, 1 M ton/y and some 1500 M EUR/y. High permeability GO (HGO, 3° dispersion) steels can achieve even higher properties. Since very large domains are detrimental, scribing the sheet surface (a series of parallel lines arrayed perpendicular to RD, spaced a few mm apart, obtained through mechanical scratching or laser irradiation) allows multiplication of domains oriented along the [001] axis in HGO sheets [14]. Surface coatings capable of exerting a tensile stress of 2 MPa to 10 MPa improve magnetic performance further.

The industrial processing of oriented silicon steels includes several steps [23]: starting from hot-rolled strip, surface descaling is performed in shot-blasting and pickling lines. Heavy cold rolling (CR) is followed by an intermediate annealing and final cold rolling (strips may be cold rolled twice on special cold rolling mills with an intermediate annealing in a continuous annealing furnace). Total CR exceeds 50%. A decarburization step is performed on strips that are coated with an annealing separator (magnesium oxide) to prevent the windings of the coiled strips from adhering to each other during subsequent high-temperature annealing.

A further high-temperature box annealing at a $T$ up to 1200°C for a treatment lasting 5 d to 7 d under a protective atmosphere, allows Goss-texture to be developed. Individual grains up to 5 mm and 20 mm can be grown. Material, further refined by diffusion annealing, is stress relief annealed and coated with insulation and thermally flattened. Stress relief annealing is performed in continuous annealing furnaces. Final steps are side trimming and slitting.

Power losses due to eddy currents can be minimized by reducing the sheet thickness $d$. Indeed, the 'classical' eddy current power losses are proportional to $d^2$ for a constant permeability and complete flux penetration. Nevertheless, for very thin sheets the domain wall spacing is reduced. The presence of concentrated electric fields at domain boundaries and of supplementary domain structure in thin sheets, limit the beneficial effect of reducing sheet thickness under a few tenths of mm [24,25]. For this reason, the industrial thinner sheets usually have $d = 0.23$ mm.

## 2.7 Fe-Ni alloys

This family of alloys, before their use as magnetic materials, was already known for the low thermal expansion coefficient of Invar, an Fe-36%Ni alloy, discovered by C. E. Guillaume in 1896 [26]. The ferromagnetic $\gamma$ phase can be retained by a suitable choice of annealing $T$, cooling rates, addition of other alloying elements such as Mo, Cu, Cr. For Ni contents above 35%, the $\gamma \rightarrow \alpha$ transition still exists, but it occurs at $T < 500$ °C and is therefore limited because of low diffusion rates (Fig. 6). Two main families of alloys for magnetic applications have been developed: the 'low-Ni' FeNi alloys, containing 47% to 50% of Ni, featuring higher saturation fields (1.6 T) and maximum permeabilities up to 60 000, and the 'high Ni' alloys including mumetal and containing approximately 80% of Ni, with lower saturation fields (0.8 T) but higher maximum permeability (up to 800 000).

High-Ni alloys were the first to find commercial applications: Elmen's work in 1913 was aimed at finding a material superior to Si-steels for 'use in telephone apparatus operating at less than a few hundred gausses'. Low-Ni alloys are prepared as strongly textured (110)[001] sheets by means of severe CR and annealing at $T \approx 1000$ °C. Magnetic annealing is usually applied. These alloys, and in particular 45 Permalloy, also found early applications in telephone apparatus, since they featured higher saturation than any of the other permalloys and could be operated at higher induction.

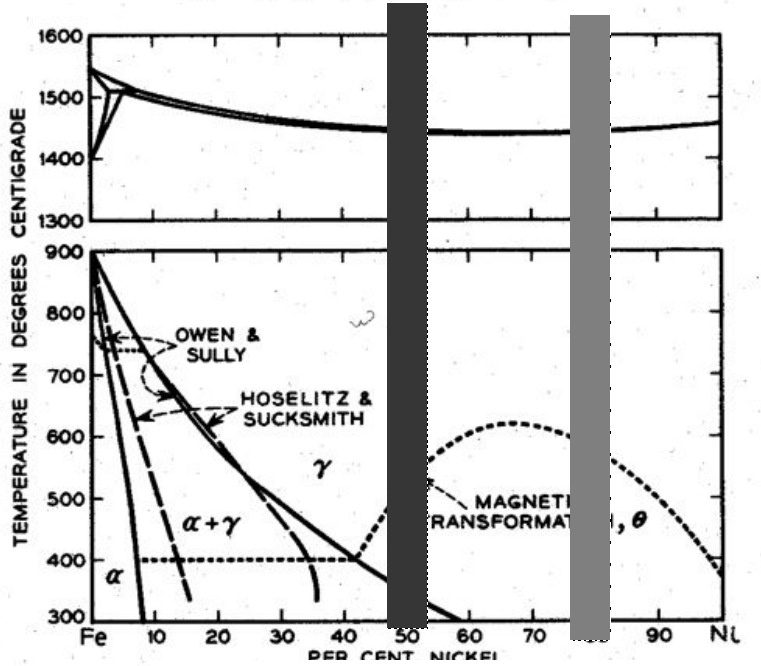

**Fig. 6:** Fe-Ni phase diagram (from Bozorth [3]), showing the two main compositions of alloys of this family for magnetic applications

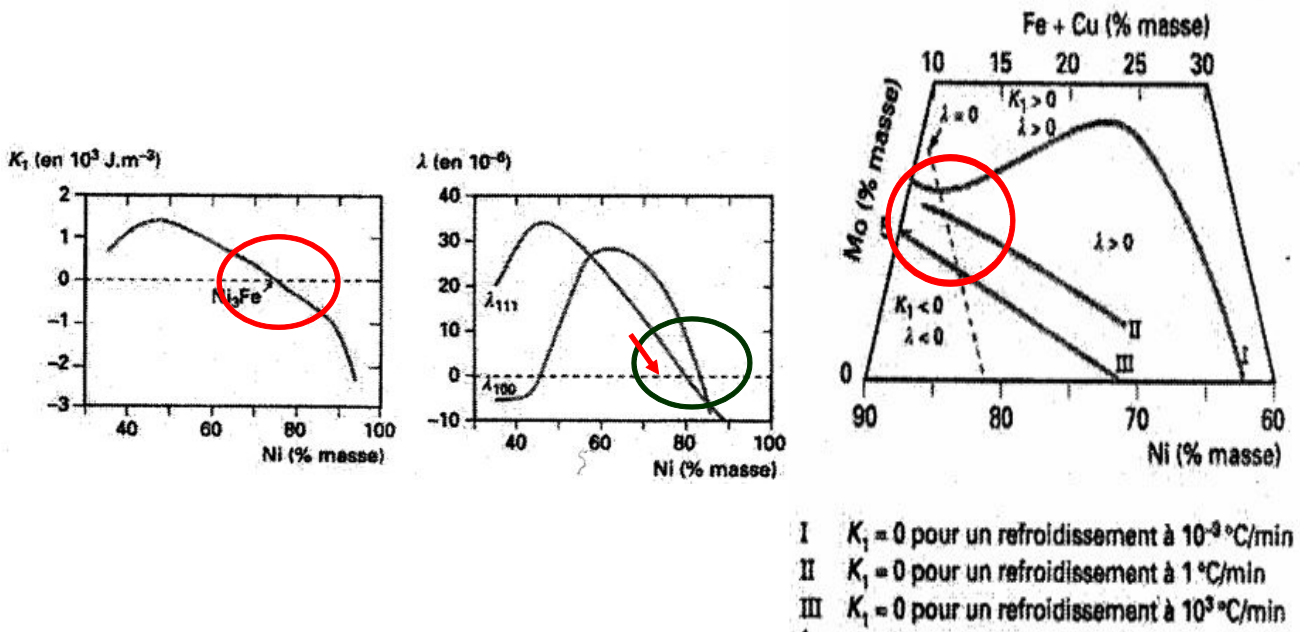

**Fig. 7:** Between 65% and 85% Ni crystalline anisotropy $K_1$ (the relevant coefficient of the first order expansion of the crystal anisotropy energy for a given angle of misorientation [6]) vanishes and magnetostriction $\lambda$ as well, but not simultaneously. Addition of other elements (Mo, Cu, Cr, etc.) allows the two parameters to be reduced simultaneously and a very high permeability to be achieved (from Ref. [27]).

As already mentioned, the highest permeability in this alloy family is featured by alloys containing approximately 80% of Ni. Indeed, between 65% and 85% Ni crystalline anysotropy vanishes and magnetostriction as well (but not simultaneously, see Fig. 7). The addition of other elements (Mo, Cu, Cr, etc.) reduces simultaneously the two parameters, while increasing the resistivity (relevant for ac applications).

Vacuum chambers for the circulating beams in the LHC injection and extraction septa [28] were manufactured from a 77Ni-5Cu-4Mo-Fe mumetal [29] produced by Imphy /FR. Because of very limited solubility of N in mumetal, mumetal tubes should be welded to AISI 304L or AISI 316L flanges depending on the application, in order to avoid gross porosity unavoidable in mumetal/AISI 316LN autogeneous welds [30].

Internal stresses, plastic deformation induced during forming and welding all degrade the magnetic performance of mumetal (Fig. 8). A high-temperature annealing treatment allows restoration of magnetic properties. Nevertheless, in order to achieve the best magnetic properties of mumetal (max. permeability as high as 100 000, coercivity as low as 0.05 Oe), the final annealing treatment should be performed under hydrogen at 1120°C.

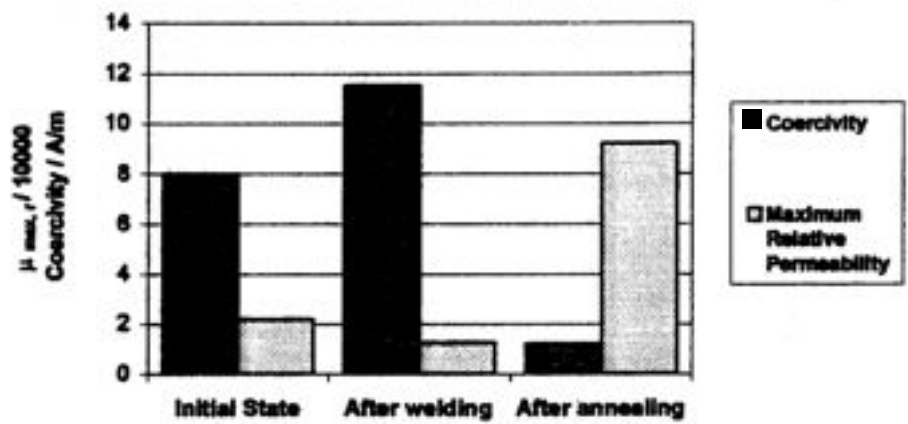

**Fig. 8:** Evolution of coercivity and maximum permeability of mumetal with fabrication steps. Sample machined from the as-delivered material ('initial state'), sample containing TIG welds ('after welding'), idem after annealing under vacuum at 1070°C for 1 h ('after annealing'). Coercivity, which was 8.0 A/m in the initial state, increased due to the presence of the welds (11.5 A/m) and was considerably reduced by annealing (1.2 A/m). Permeability, approximately 21 800 in the as delivered state, decreased about 60% due to the presence of the welds, to a value of around 12 570. After the annealing treatment, the permeability reached 92 000 [30]. Material delivered by Telcon Ltd. /UK.

For some Fe-Ni alloys, particular care has to be paid to cooling conditions from an 'ordering temperature' ($\cong 500°C$). Heat treatment in a magnetic field, discovered by Kelsall in 1934 [31] can cause a large increase in permeability ($\times 10$) of some Fe-Ni alloys. Taking as an example 65 Permalloy, the presence of the field during cooling from $T_c$ to 400°C is essential. Magnetic annealing is based on non-random diffusion of atoms and preferential alignment of like-atom pairs. It is explained on the basis of a 'directional-order' theory [32–34]. At a temperature $T \leq T_c$, but high enough for diffusion to occur, like-atom pairs tend to be aligned in the direction of the local magnetization. As temperature is lowered during cooling, since diffusion constants become too low for further diffusion to occur, the 'freezing' in place of like-atom pairs produces a uniaxial anisotropy in the material [35].

## 2.8 Compressed powdered iron and iron alloys

Iron-based products produced by powder metallurgy (PM) techniques have the advantage of featuring isotropic 3-D properties. For application in the medium frequency (kHz) range, cores of compacted iron are often used. PM products can be near-net shaped to close dimensional tolerances and show satisfactory $T$ stability due to limited internal stresses. Designers can exploit 3-D flux paths [36]. For some components (claw pole, brushless motor) use of laminations would not be applicable.

Coated iron powders (of a typical size 50 μm to 100 μm) are mixed with some 1% of binding material, compressed as cores of the desired shape and then sintered. The cores are coated by protective painting. Fe, Fe-Si and other Fe powders can be compacted. The maximum achievable permeability is controlled by grain size, sintering $T$ and degree of porosity. The dc properties of hot-pressed high-purity Fe are considered as good as or better than conventional Fe. Electromagnetic actuators of complex core shapes for use in transport, electrical rotating machines are produced through this technology. The usual field of compressed powdered irons are dc and medium frequency applications (ferrites dominate in very high frequencies). Two components of core losses are identified in powder metallurgy irons: a classical hysteresis contribution and an eddy current one [37]. Powder metallurgy cores exhibit larger hysteresis contributions than steel sheets, but (particle to particle) eddy current contributions are much smaller already at 60 Hz (estimated at 5% of the total in the example of Fig. 8). Eddy currents become dominant at higher frequencies, where, due to lower eddy current contributions, pressed materials are competitive in performance with laminated steels (Fig. 9). In powder materials, two kinds of eddy current are identified, circulating within the insulating particles and around clusters of particles. This explains size effect contributions to the total losses [38], and the advantage of the use of insulating coatings on powder particles, respectively.

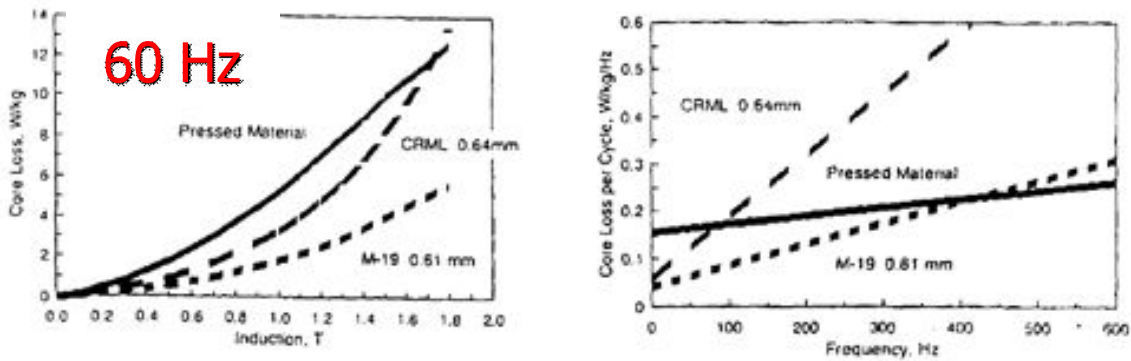

**Fig. 9:** Comparison of a pressed material with 0.64 mm thick 'Cold Rolled Motor Lamination' (CRML) steel and 0.061 mm thick 'M-19' NGO silicon steel. At 60 Hz, the total core loss of the pressed material is dominated by hysteresis contributions, while the eddy current contribution is estimated at only 5% of the total losses. The pressed material has lower performance than conventional steels. At higher frequencies, the total loss for the pressed material is less than for laminated steels (from Ref. [37]).

## 2.9 Soft spinel ferrites

Ferrites are largely applied in the high-frequency range up to few hundred MHz. Their composition is $MO \cdot Fe_2O_3$ where M is a divalent metal (M = $Fe^{2+}$, $Mg^{2+}$, $Mn^{2+}$, $Ni^{2+}$, $Zn^{2+}$; lodestone $FeO \cdot Fe_2O_3 = Fe_3O_4$ is a particular case of a ferrite). They are ceramics featuring very high resistivity, between $10^6$ Ω m and $10^{12}$ Ω m. They are used at frequencies where eddy current losses for metals become excessive. At very high frequencies, they are ideal soft magnetic materials. Disadvantages are low magnetic saturation (typical range 0.15 T to 0.6 T), low $T_c$ (330°C to 585°C) poor mechanical properties, hardness and brittleness. Since they are practically unmachinable, close dimensional tolerances are achieved by grinding.

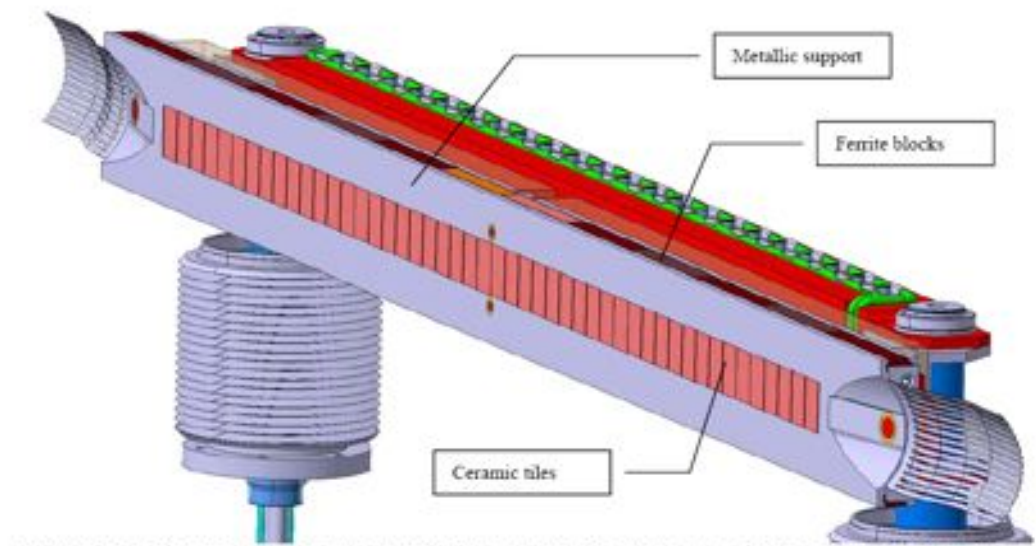

**Fig. 10:** Collimator jaw equipped with BPM buttons and ferrite blocks (courtesy of A. Dallocchio)

Ferrites are used as cores for electromagnetic interference suppression, to control transmission or adsorption of electromagnetic waves. One example of the application of ferrites in accelerators is the LHC collimators, where ferrite blocks are used to damp trapped modes (Fig. 10).

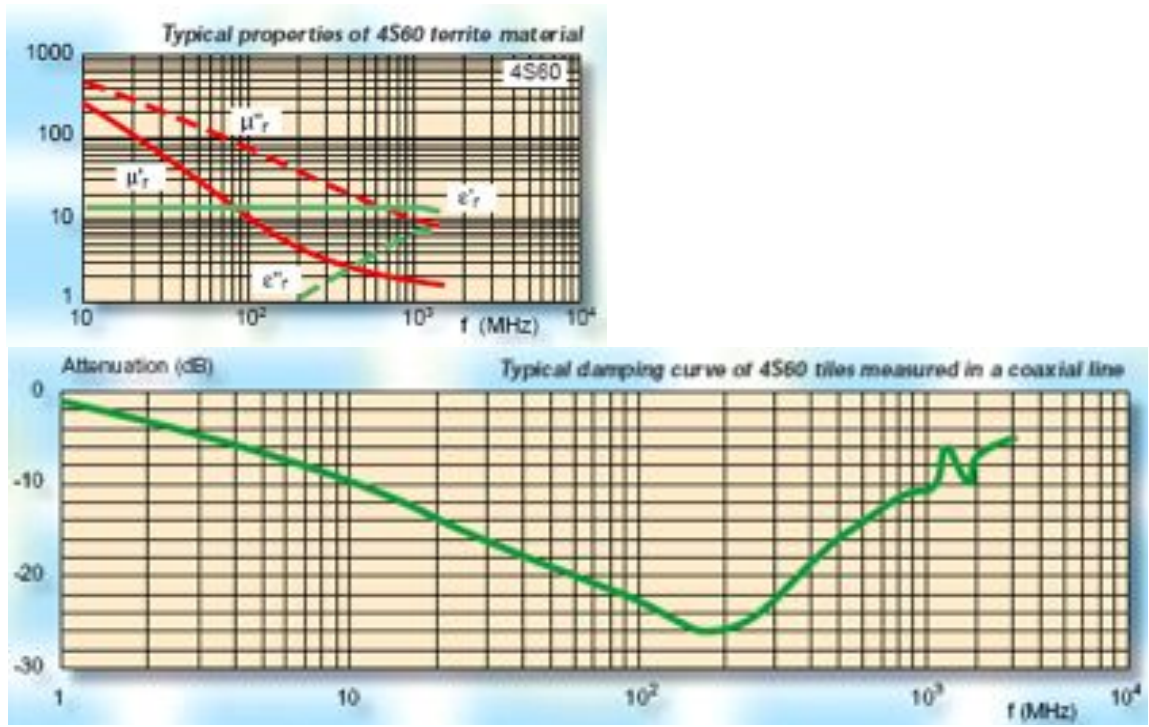

**Fig. 11:** Properties of FERROXCUBE 4S60 ferrite. Maximum absorption occurs where $\mu_r$ matches the dielectric constant $\varepsilon_r$ (from Ref. [39])

The ferrite FERROXCUBE 4S60 is used for the TCLIA collimators in the LHC at CERN. This 'NiZn' ferrite is of a mixed type. The proportions of Ni and Zn are tailored for the specific application. While MnZn ferrites show the highest permeability, NiZn, shows a broadband operation up to 1000 MHz and features higher resistivity (Fig. 11).

Ferrites are prepared through a powder metallurgy process. After mixing and weighing the base oxides in the form of fine powders, the powder is heated to between 900°C and 1200°C (pre-firing in air) to produce flakes of a few cm$^3$. During this stage, the spinel structure is formed by the reaction of $Fe_2O_3$ with MO. This material is ball milled in water and mixed with a binder. A spray drying step results in balls of few mm$^3$. The powder is then mechanically pressed in a mould by die punching or hydrostatic pressing. The green (filling factor 50% to 60%) is batch sintered in a kiln at $T$ between 1200°C and 1400°C, with or without external pressure, in an oxidizing atmosphere (filling factor 95% to 98%). Shape is conferred by a final grinding [14,40].

## 2.10 Innovative materials: amorphous alloys

Amorphous alloys (metallic glasses) are materials devoid of long-range atomic order. They are produced by rapid solidification from the liquid or gaseous state. Generally they are in ribbon form and obtained by rapid solidification from the liquid state. Metastable sputtered amorphous thin-films of Co-Au were produced by Mader and Nowick in 1965 [41]. Today, ribbons up to 100 mm or 200 mm wide of a thickness of 10 μm to 40 μm can be produced. In order to develop the amorphous structure, cooling rates of $10^5$°C/s to $10^6$°C/s have to be obtained. Amorphous wires are also produced. The preparation of amorphous magnetic materials by rapid quenching from the melt is generally performed through planar flow casting on metallic wheels or drums in air, shielding with atmosphere or vacuum (Fig. 12).

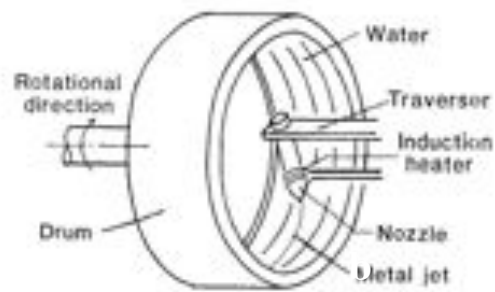

**Fig. 12:** Rotating drum. A jet of molten metal is ejected into a rotating water layer (from Ref. [42]). In order to guarantee rapid quenching, drum velocities are as high as 3 m/s to 10 m/s. Reported range of peripheral velocities of metallic wheels for the production is 10 m/s to 40 m/s [14].

Amorphous alloys are generally of the type $T_{70-80}M_{30-20}$, where T is a transition metal (Fe, Co, Ni) and M is a combination of metalloids (B, Si, P, C, etc.). Amorphous ribbons of 25 μm thickness in $Fe_{78}B_{13}Si_9$, compared to a conventional GO Fe-3Si of 0.23 mm of thickness, show a coercive field of 2 A/m after annealing (5 A/m for the latter) and a maximum relative permeability of $2 \cdot 10^5$ ($8 \cdot 10^4$, idem). Annealing allows for a local crystallization. These materials are hard and show ductility limited to some 2.5% elongation at failure. Alloy glass ribbons have an excellent magnetic softness. Magnetization curves have an ideal loop with a precise 'cut' at a specific value of $B$ (see the example of the loop of a commercial Vitrovac material in Fig. 13a). This behaviour, combined with a high magnetomechanical coupling, makes them ideal materials for sensors for electronic article surveillance in the so-called 'harmonic-electromagnetic systems': glass ribbons are used for security tags in libraries and stores [43, 44]. In order to be conveniently activable and deactivable and not to be demagnetized by the fields in the interrogation zone, materials should show $H_c$ between 1600 A/m and

8000 A/m [43]. Permalloys could also be used for the purpose illustrated in Fig. 13, but their ductility would imply a risk of degradation of their magnetic properties by handling.

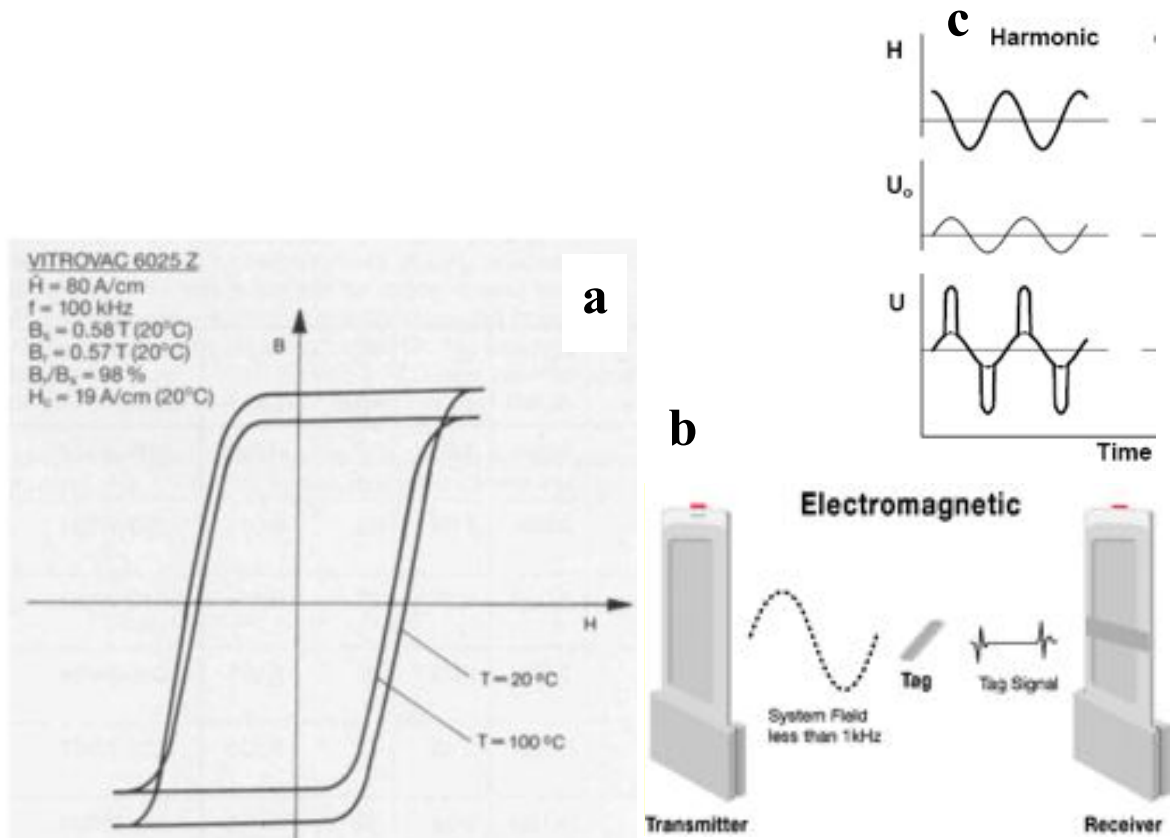

**Fig. 13:** a) Ideal loop of a commercial amorphous material (from Vacuumschmelze [45]). When applied in security tags, a length of amorphous alloy is packaged together with a hard magnet strip. The tag is activated by unmagnetizing the permanent magnet, and deactivated at the cash register by magnetizing (saturating) the strip. b) Activated labels respond to an electromagnetic field generated by a pedestal at the store exit with specific frequencies detected by the receiver, since the material is periodically driven into saturation. c) Response to a periodic excitation $H$: $U_0$, unactivated label, $U$ activated label: high permeability at fields below saturation induces high attenuation when the label is activated, while above saturation spikes are present during the transmission time (from G. Herzer [43]).

### 2.11 Innovative materials: nanocrystalline alloys

Nanocrystalline alloys consist of small ferromagnetic crystallites of bcc FeSi with grains of 10 nm to 15 nm embedded in an amorphous matrix, coupled to each other by exchange interaction. Crystallites are separated by 1–2 nm for interaction. These materials show very low coercivity (0.4 A/m to 8 A/m), high initial permeability (up to 150 000), low losses and magnetostriction and high saturation up to 1.3 T. They are available in ribbons of few tens of μm thickness. Compared to conventional Fe-Si steels and amorphous materials, they show exceptionally low core losses, specially at very high magnetization rates (Fig. 14). For this reason, they are envisaged for application to heavy-ion inertial fusion-energy, based on induction accelerators where some 30 000 tons of magnetic material are necessary for induction cores [46]. For this application, nanocrystalline materials would show the best performance in terms of core losses at the required magnetization rates between $10^5$ T/s and $10^7$ T/s.

They are commercially applied in low-loss high-frequency transformers, such as the one foreseen by the Linac group of the Spallation Neutron Source (SNS) of the Los Alamos National Laboratory for application in high-frequency polyphase resonant converters for the ILC (International Linear Collider) [47]. These materials are interesting for both high-energy physics and accelerator applications, but a drawback is their cost (between $20 and $150 per kilogram).

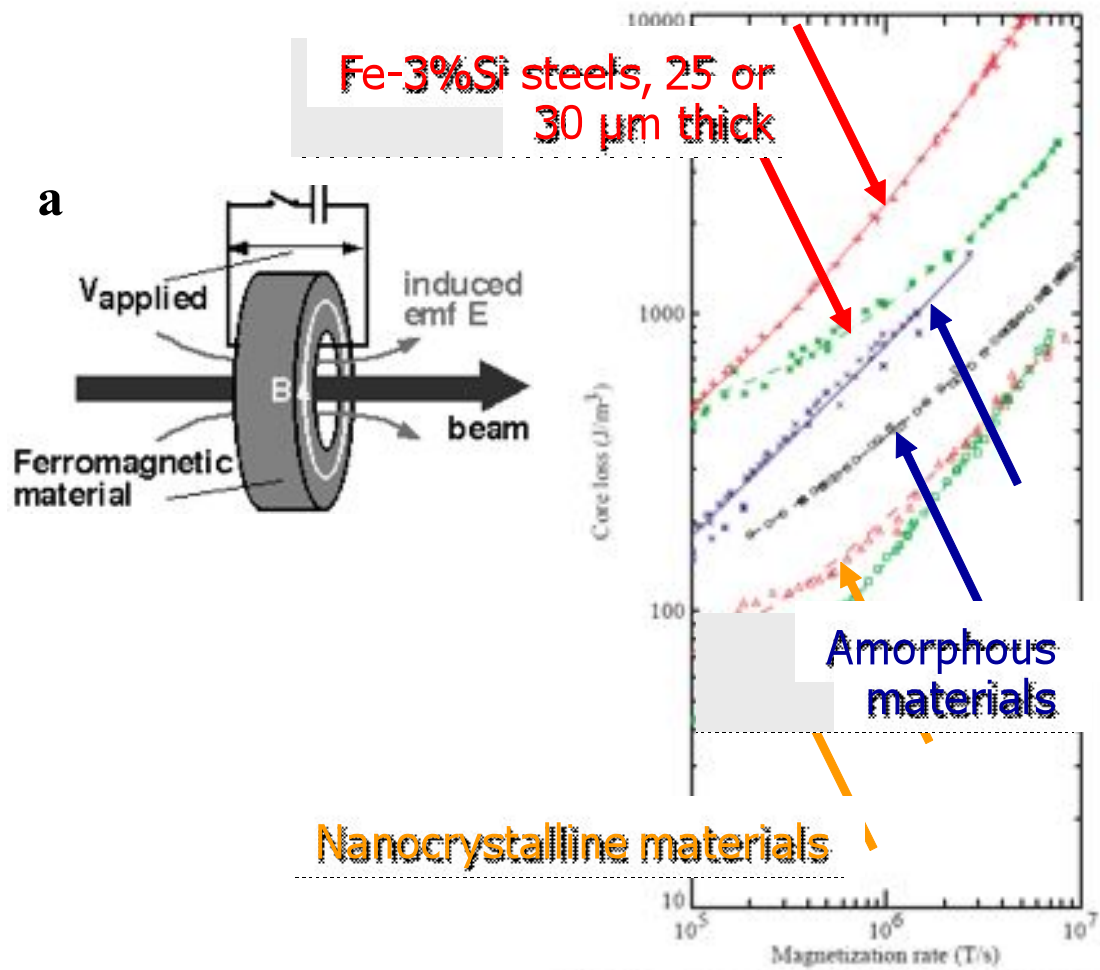

**Fig. 14:** a) In heavy-ion inertial fusion-energy accelerators some 30 000 tons of magnetic material are necessary for induction cores in order to accelerate heavy-ion energies in the GeV range and deliver several MJ per pulse to a target (from Ref. [46]). b) Nanocrystalline materials are the highest performing ferromagnetic materials as induction core alloys. They form the lowest loss group at the very high magnetization rates required for this application (from Ref. [48]).

## 3 Methods of measurement

### 3.1 Characterization of soft magnetic materials

Characterization of the magnetic properties of soft magnetic materials is generally based on the measurement of a transient voltage induced on a secondary winding by a step-like field variation applied on a primary winding. The two coils are generally wound together on a toroid sample. The signal is integrated over a time interval for complete decay of eddy currents, since every recorded point should correspond to a stable microscopic configuration of the system. Since flux variations are measured, providing a discrete sequence of field values, a reference condition is needed, that is generally the saturated or demagnetized state. A 'split-coil' permeameter (Fig. 15) is available at CERN, facilitating the exchange of samples, and not requiring a prior winding contrary to the wound toroid methods [49]. For a careful assessment of the magnetic properties through these methods, a precise measure of the cross sectional areas of the samples is necessary, which should not be estimated from sample mass and density.

The technique of the split-coil permeameter has recently been used for the electromagnetic characterization of the steels used for the OPERA magnets [50].

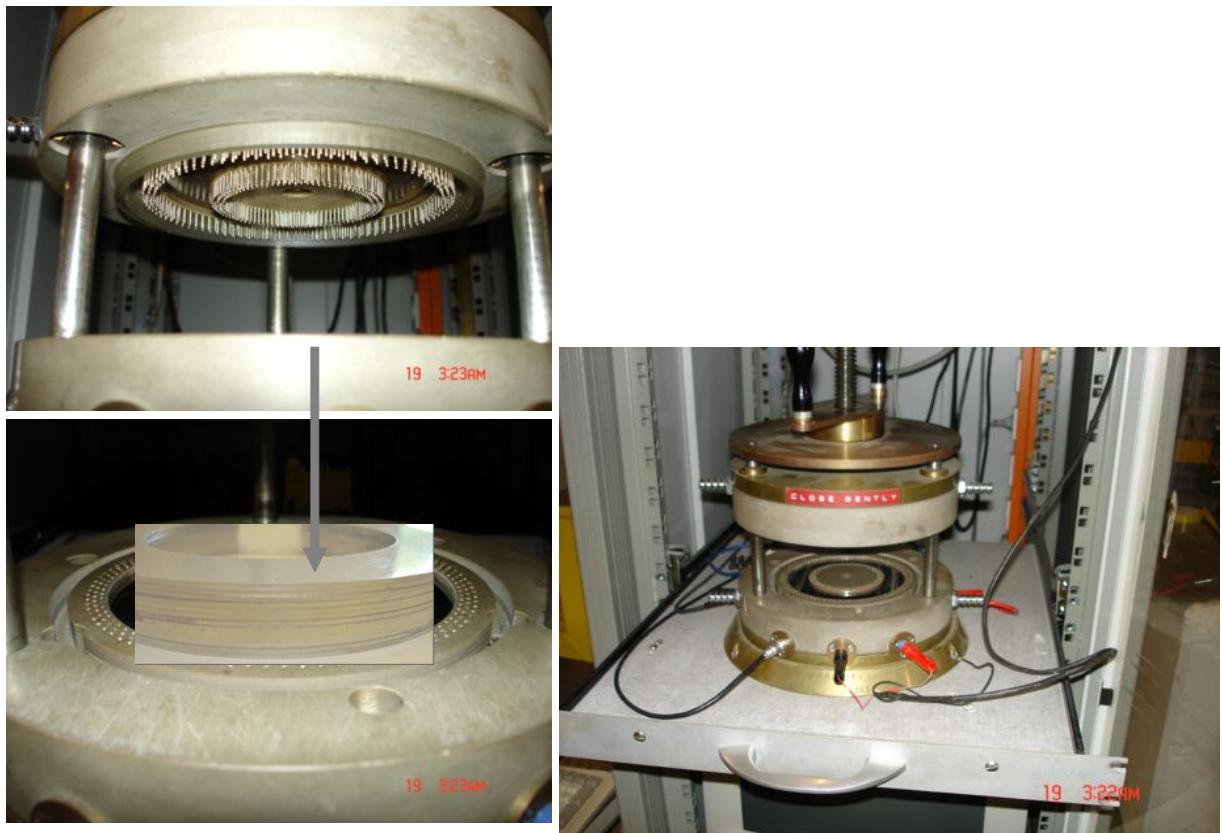

**Fig. 15:** Split-coil permeameter. The flux measuring and excitation coils are not wound directly onto the samples. The advantage of the system is the rapid exchange of samples (rings, or pile up of rings), an automatic measurement of the relevant parameters of the hysteresis curve, making it adapted to the evaluation of large series. A drawback is the cumulative contact resistance between the two split parts of the coils (two contacts per turn), making this method inadequate for measurements in liquid He (excessive power dissipation).

CERN owns special coercimeters aimed at measuring the coercivity directly on steel sheets without having to cut samples. The instrument was used to perform the coercivity measurements needed during the production of the 11000 t of steel sheets for the LEP dipole magnets and to measure their permeability. The coercimeter is based on a main excitation and detector coil, and auxiliary coils used to estimate the air gap of the yoke contacts (Fig. 16). The yokes are pressed against the sheets that are introduced through a system of rollers in the coercimeter. The sequence of measurements is automatic. After demagnetizing the yokes, a stable hysteresis cycle is assessed. Four flux variations

$Δφ_i$ ($i$ = 1 to 4) are measured along the cycle and the remanent flux $φ_r$ is calculated as $φ_r = ¼ [|Δφ_1| - |Δφ_2| + |Δφ_3| - |Δφ_4|]$. The current $I_c$, necessary to cancel $φ_r$ is measured on both sides of the cycle, allowing the coercive field to be calculated from the mean absolute value of $I_c$ taking also into account the coercive field of the yoke (from Ref. [51]).

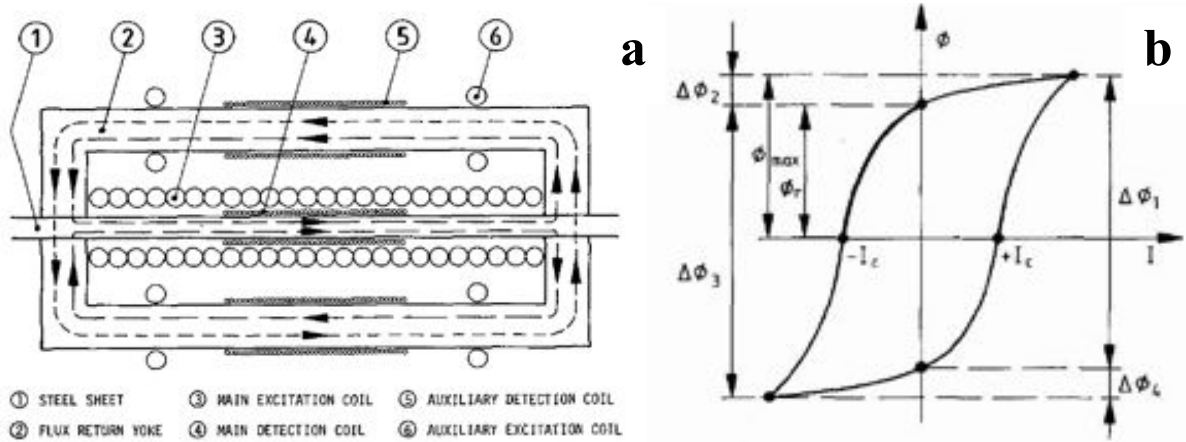

**Fig. 16:** a) The coercimeter is based on a main excitation and detector coil. b) hysteresis cycle (from Ref. [51])

## 3.2 Measurement of feebly magnetic materials

Measurement of permeability of feebly magnetic materials (materials with permeability in the range $μ_r$ = 1.00001 to 2 or 4) is possible through several techniques. Portable magnetometers are adapted to non-destructive measurement of materials of complex shape but having a radius of curvature not less than 40 mm or a flat area not less than 20 mm in diameter. The material should be thicker than 8 mm. Measurements of materials thinner than 8 mm can be performed by stacking two pieces or applying corrections. The air gap between two pieces should be as small as possible; otherwise permeability less than actual value will be estimated.

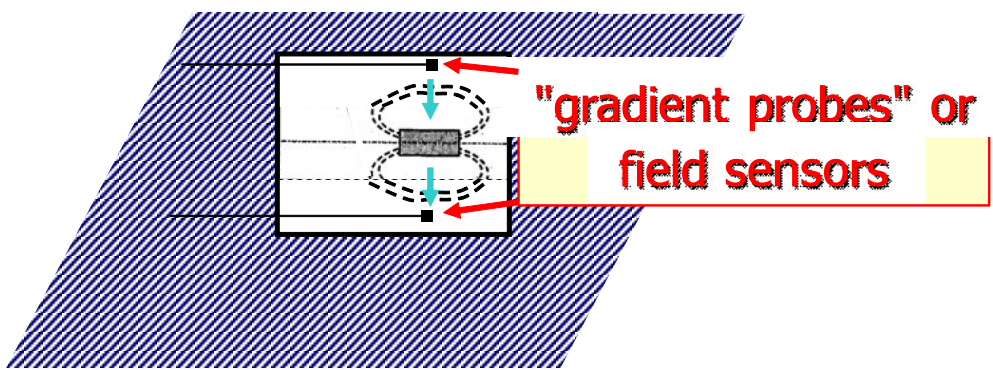

**Fig. 17:** Gradient probes allow measurement of the displacement of the zero field line of the permanent magnet towards the feebly magnetic material to be measured. The field strength of the probes generally used to assess permeability of austenitic stainless steels is approximately 80 kA/m.

Probes of portable magnetometers are based on a permanent cylindrical magnet (Fig. 17). A permanently built-in field sensor, sensitive to the field emanated from the sample, is placed on either side of the cylindrical magnet in the plane perpendicular to the cylinder axis at the centre of the permanent magnet. Since the permeability of the feebly magnetic material to be measured is larger than 1, the zero field line of the cylindrical magnet will be displaced towards the sample. This displacement allows the permeability of the material to be assessed [52].

Magnetic balance measurements are suitable for measurement of materials with $\mu < 1.05$. This destructive test method is applicable to the evaluation of semifinished products or welds before fabrication of parts. Samples measurable at CERN are cylinders with a diameter of 3 mm and a height of 2 mm. Larger samples are measurable according to relevant standards. After several reverses of the current to delete the effects of the hysteresis in the core, an increasing current is turned on in an electromagnet, producing step-by-step increasing field strength up to more than 450 kA/m. The sample should not be overheated. The sample is suspended from the balance and positioned above the centreline of the air gap of the electromagnet (Fig. 18). The permeability is calculated from the apparent change in mass of the specimen [53].

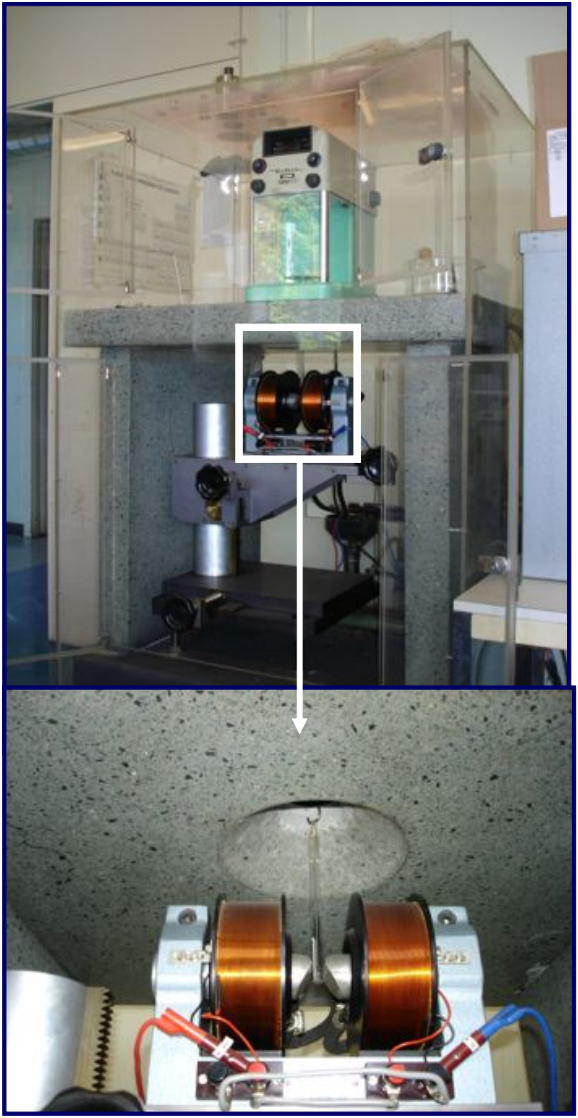
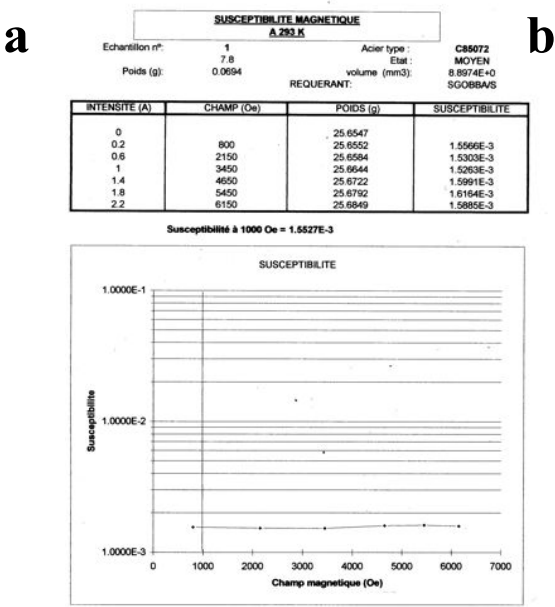

**Fig. 18:** a) Magnetic balance built up at CERN. b) Magnetic permeability is assessed for different fields from the apparent change of mass of the sample.

Magnetic Field Microscopy (MFM) techniques allow identification of a magnetic contrast between magnetic and non-magnetic phases on a very local scale (Fig. 19). More recently, MOKE (Magneto-Optical Kerr Effect) techniques have been associated to MFM and used to quantify locally the magnetic properties of phases in the samples, including measurement of hysteresis loops [54].

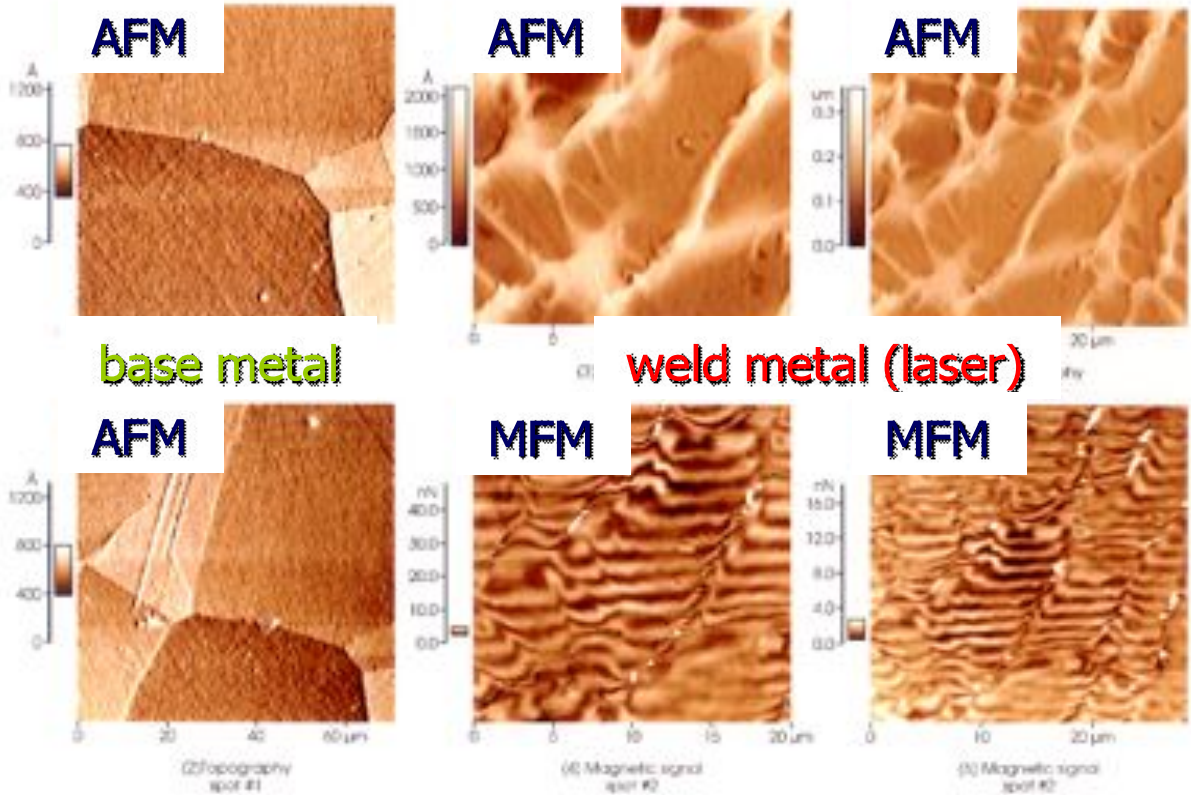

**Fig. 19:** Atomic Force Microscopy (AFM) and MFM measurements of a sample of an austenitic stainless steel including a longitudinal laser weld, after tensile testing at 4.2 K. While AFM images are only sensitive to the topographical contrast of the surface (grain boundaries and dendrite boundaries are visible in AFM images of the base metal and the weld, respectively), MFM allows qualitative identification of the presence of a magnetic contrast between magnetic and non-magnetic phases possibly present in the weld.

## 4 Magnetic lag

The two main sources of magnetic lag are discussed: lag due to eddy currents, and the so-called magnetic 'after-effect' (Nachwirkung), which is material dependent.

### 4.1 Lag due to eddy currents

When one applies a magnetizing current to a bar, eddy currents develop whose direction is opposite to that of the applied current. This current flow has both an effect in DC (the field cannot penetrate immediately into the interior of the material) and in AC (set-up of field gradient between surface and interior). Lag due to eddy currents should be taken into account when measuring the magnetization curves of soft magnetic materials (see Section 3.1). Indeed, the sudden application of a field to a cylinder of field-dependent permeability of diametre $d$ requires a time $\tau$ to reach a field $B$:

$$\tau = 0.55 \frac{d^2}{\rho} \int_b^1 dB/dH \frac{db}{b} \qquad (5)$$

where $b = 1 - B/B_0$, $B_0$ is the ultimate field and $\rho$ is the resistivity [3]. For a constant permeability, a variation of field applied to a lamination of relative permeability $\mu_r$, conductivity $\sigma$, thickness $d$ implies a decay time of eddy-current-generated counterfield $\tau \propto \mu_0 \mu_r \sigma d^2$ [14]. For a 10 mm sheet of a 1010 steel, $\tau$ is approximately 3 s [55].

Undesirable effects due to lag induced by eddy currents were identified in CERN SPS magnets, when measuring in multicycles the influence of the 450 GeV proton cycle on the following positron cycle [56].

## 4.2 Magnetic 'after-effect'

The application of a magnetic field requires a given time to reach the final induction value in a magnetic material. Apart from the contribution to the lag due to eddy currents discussed in Section 4.1, additional delayed effects have a metallurgical origin (impurities such as C and N in Fe, dislocations, etc.) and are mainly due to a time-dependent microstructural redistribution associated with strain induced by magnetostrictive effects. The commonalty of phenomena between magnetic (Fig. 20a) and anelastic (Fig. 20b) after-effect suggests a common origin. The time constant of magnetic after-effects can be appreciable. The effect is strongly dependent on temperature (faster for higher temperatures) and on material purity. A detailed discussion of the different contributions to magnetic after-effect is found in Ref. [57].

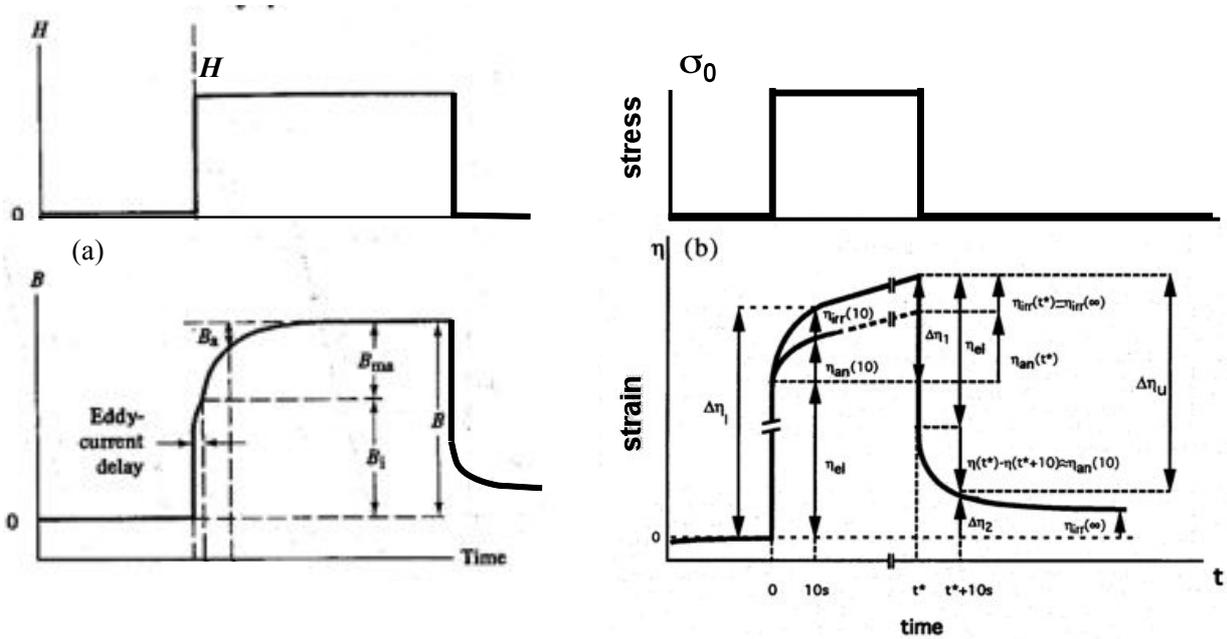

**Fig. 20:** Analogy between magnetic (a, from Ref. [6]) and anelastic (b, from Refs. [58, 59]) after effect. In Fig. 20a, the field strength in a demagnetized sample is increased suddenly from zero to a constant value $H$. The induction $B$, apart from the delay due to eddy currents, rises immediately to a value $B_i$, and then at a finite rate to the final value $B$ that is associated to $H$ in the magnetization curve. In Fig. 20b, the sudden application of a stress $\sigma_0$ to a sample suddenly induces an elastic strain $\eta_{el}$, followed by an elastic after-effect (time-delayed deformation), composed of reversible anelastic ($\eta_{an}$) and irreversible viscoelastic ($\eta_{irr}$) contributions. Delayed effects are observed as well on removal of the field (magnetic or elastic).

# 5 Conclusions

Magnetic materials are key elements of magnet technology. They should be procured on the basis of careful selection and adapted specifications, since their primary and secondary metallurgy, chemical composition, purity, applied thermal treatments, and microstructure will have a significant influence on their final properties. Low-carbon steel laminations, but also general-purpose constructional steels, such as type 1010, generally used for applications that require 'less than superior' magnetic properties [4], are often applied as yoke materials for accelerator and experiment magnets. They are not always purchased to magnetic specifications. Soft ferromagnetic materials of better controlled composition and impurity limits, properties and metallurgy might be considered for specific applications, such as fast magnet systems. On the other hand, innovative materials such as nanocrystalline and amorphous alloys are being considered or are already used for an increasing number of devices, including for high-energy physics and fusion-related applications. Examples are high-frequency transformers for the International Linear Collider and induction cores of heavy-ion inertial fusion-energy accelerators, respectively. In 2007 nanocrystalline materials represented a production of 1000 t. The importance and use of powder metallurgy is also increasing for application to structural components of magnets, soft magnetic materials, and materials for permanent magnets. In 2003, powder-based soft ferrites represented 5% of the world market of magnetic materials, including semihard and hard materials, compared to 27% covered by conventional steels [60].

**Acknowledgement**

The author wishes to thank L. Walckiers and D. Brandt for their invitation to contribute to the special CERN Accelerator School, D. Tommasini for fruitful exchanges, G. Peiro and J. Garcia Perez for helpful discussion on testing facilities available at CERN, F. Caspers for the expertise provided on soft ferrites, and R. Veness for the final spell checking of the manuscript.